\documentclass[aps,prl,amssymb,amsmath,superscriptaddress,twocolumn,10pt]{revtex4-1}
\usepackage{color, graphicx}

\def\bra#1{\langle{#1}|}
\def\ket#1{|{#1}\rangle}

\def\ee{{\mathrm{e}}}
\def\ii{{\mathrm{i}}}

\newcommand{\As}{a}

\def\bra#1{\langle #1 |}
\def\ket#1{| #1 \rangle}

\def\Ord{\mathrm{O}}

\renewcommand{\Im}{\mathrm{\,Im\,}}

\def\ii{\mathrm{i}}

\newcommand{\unibari}{Dipartimento di Fisica and MECENAS - Universit\`{a} di Bari, I-70126 Bari, Italy}
\newcommand{\infn}{INFN, Sezione di Bari, I-70126 Bari, Italy}

\newcommand{\polimi}{Dipartimento di Fisica - Politecnico di Milano, I-20133 Milano, Italy}
\newcommand{\ifn}{Istituto di Fotonica e Nanotecnologie - Consiglio Nazionale delle Ricerche, I-20133 Milano, Italy}
\newcommand{\uniroma}{Dipartimento di Fisica - Sapienza Universit\`{a} di Roma, I-00185 Roma, Italy} 
\newcommand{\waseda}{Department of Physics, Waseda University, 169-8555 Tokyo, Japan}

\begin{document}

\title{Experimental investigation of quantum decay at short,\\ intermediate and long times via integrated photonics}

\author{Andrea Crespi}
\affiliation{\polimi}
\affiliation{\ifn}

\author{Francesco V. Pepe}
\affiliation{\infn}

\author{Paolo Facchi}
\affiliation{\unibari}
\affiliation{\infn}

\author{Fabio Sciarrino}
\affiliation{\uniroma}
\affiliation{\ifn}

\author{Paolo Mataloni}
\affiliation{\uniroma}
\affiliation{\ifn}

\author{Hiromichi Nakazato}
\affiliation{\waseda}

\author{Saverio Pascazio}
\affiliation{\unibari}
\affiliation{Istituto Nazionale di Ottica (INO-CNR), I-50125 Firenze, Italy}
\affiliation{\infn}

\author{Roberto Osellame}
\affiliation{\ifn}
\affiliation{\polimi}

\begin{abstract}
The decay of an unstable system is usually described by an exponential law.
Quantum mechanics predicts strong deviations of the survival
probability from the exponential: indeed, the decay is initially quadratic, while at very large times it follows a power law
{, with superimposed oscillations. The latter regime is particularly elusive and difficult to observe.}
Here we employ arrays of single-mode optical waveguides, fabricated by femtosecond laser direct inscription, to implement quantum systems where a discrete state is coupled and can decay into a continuum. The optical modes correspond to distinct quantum states of the photon and the temporal evolution of the quantum system is mapped into the spatial propagation coordinate. By injecting coherent light states in the fabricated photonic structures and by measuring light with an unprecedented dynamic range, we are able to experimentally observe not only the exponential decay regime, but also the quadratic Zeno region and the power-law decay at long evolution times.
\end{abstract}

\maketitle

\begin{figure}[t]
	\centering
\includegraphics[scale=0.9]{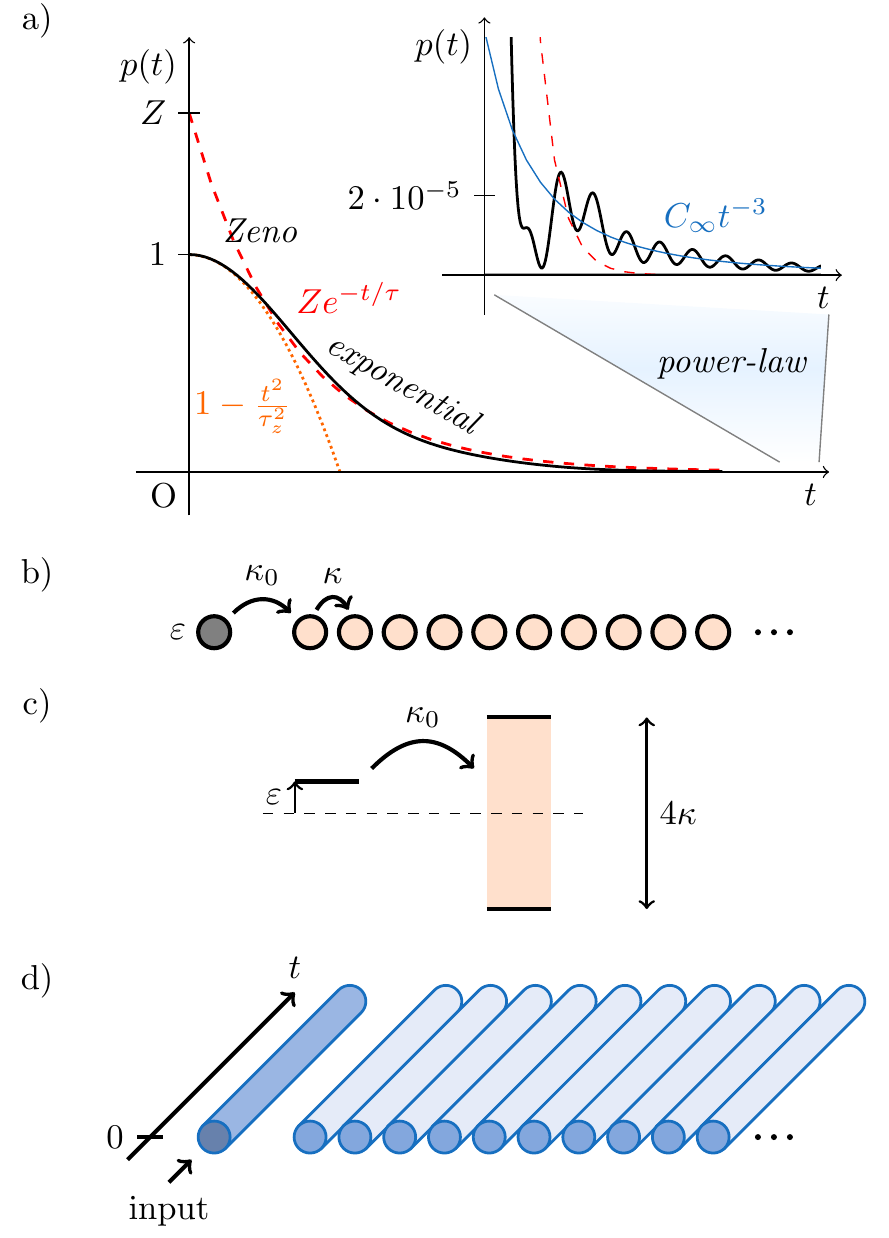}
	\caption{a) Typical decay of a quantum state coupled to a continuum, showing the peculiar features of the survival probability $p(t)$. The initial region is quadratic, with the curvature at $t=0$ characterized by the Zeno time $\tau_Z$. At intermediate times a familiar exponential decay sets in, with a lifetime $\tau=\Ord(\lambda^{-2})$, where $\lambda$ is the coupling constant, and a ``wavefunction renormalization'' $Z=1+\Ord(\lambda^2)$, which represents the value of its extrapolation back to $t=0$. At very large times, when the survival probability is reduced to $\Ord(\lambda^{10})$, a power-law regime is observed, with superimposed oscillations.
	In particular, we refer to the system represented in~(b): an optical mode with  detuning $\varepsilon$ is coupled, by a coupling constant $\kappa_0$, to a chain of optical modes with relative coupling $\kappa$. c) Level scheme of such a system: the states of the chain make up a continuum with bandwidth $4 \kappa$ {(as in a tight-binding one-dimensional lattice).} d) Experimentally, we  study this system with an array of single-mode optical waveguides. 	}
	\label{fig:genevol}
\end{figure}

The exponential decay law is commonly associated to the probability that a system, initially prepared in an unstable state (such as an excited atomic level or an unstable elementary particle), is observed in the same state after some time \cite{expon1,expon2}. Actually, for quantum mechanical unstable states, decay can only be approximately exponential \cite{strev,zenoreview1,zenoreview2}: at short times the survival probability is quadratic, while at long times it is dominated by a power law (see Fig.~\ref{fig:genevol}a). The aforementioned features of the quantum evolution are consequences of first principles and represent strong signatures of non-classical behavior. The initial quadratic behavior, also known as the Zeno regime, stems directly from a short-time expansion of the Schr\"odinger evolution, with the only hypotheses of normalizability of the wave function and finite energy fluctuations of the initial state.
The familiar exponential decay sets in at intermediate times and its derivation is always the consequence of assumptions of some sort, such as weak coupling or Markovianity.
The long-time evolution is a consequence of the boundedness from below of the Hamiltonian, an indispensable condition from a physical perspective. Under this hypothesis, a straightforward application of the Paley-Wiener theorem on Fourier transforms yields long-time power-law tails~\cite{Khalfin57,Khalfin58,Exner85} (see, however,~\cite{ref:exponential}).

{The initial ``Zeno'' region \cite{MS} has been experimentally confirmed in a variety of physical systems, including trapped atoms \cite{Wilkinson,FGR,Wineland}, Bose-Einstein condensates \cite{Ketterle,Firenze2014},  cavity quantum electrodynamics \cite{Raimond2010, Reichel}, Rydberg atoms \cite{Signoles2014} and optical waveguide arrays \cite{biagioni}. On the other hand, to the best of our knowledge, only a single experimental observation of the power-law decay was reported, with the observation of the temporal decay of the fluorescence signal of dissolved organic molecules \cite{rothe}, although without an underlying theoretical model that enables one to compute the power-law decay from first principles. In general, power-law decay tails are very elusive,} since the preceding region of exponential decay usually depletes the initial state at a point that makes any subsequent observation {extremely challenging}.

Arrays of single-mode optical waveguides are a powerful platform to experimentally investigate diverse quantum dynamics. The optical modes represent distinct quantum states of the photon, that can be coupled with high control by tuning the evanescent-field-mediated coupling between waveguides. The time evolution of the Schr\"odinger equation is mapped onto the longitudinal propagation in the waveguides, thus making it easy to investigate even fast dynamics \cite{longhiRev}. In addition, photons are almost immune to decoherence. Exploiting such favourable features, several quantum phenomena that are difficult to observe in solid state systems have been successfully studied with photonic structures. These include Bloch oscillations \cite{chiodo, trompeter}, Anderson localization \cite{lahini, martin} and the Zeno decay regime mentioned above \cite{biagioni}. Engineered waveguide arrays, excited with identical photon pairs, have also allowed the experimental study of multi-particle quantum decay processes \cite{crespiFano}. 

In this work, we use optical waveguide arrays, fabricated by the femtosecond laser micromachining technology \cite{szameitNolte,chiodo,crespiFano}, to implement quantum systems where a discrete state is made unstable by its coupling to a continuum \cite{cohentannoudji}. Different structures are fabricated, optimizing  the parameters in such a way that different dynamical decay regimes can be detected and scrutinized, when coherent laser light is injected at the input. To probe the system evolution we acquire with high dynamic range the light scattered from the array. In this way we are able to observe, within a single experimental platform, the quadratic Zeno region, the transition to the exponential regime, the wave-function renormalization and the power-law decay at long evolution times.

The physical system investigated consists of a semi-infinite linear array of single mode optical waveguides, which can be excited by light with fixed polarization. The transverse optical modes correspond to localized quantum states $\ket{n}$, with $n \geq 0$ indexing the different waveguides.
Neighboring modes are coupled by evanescent-field interaction: the first one is coupled to the second one by a coefficient $\kappa_0$, while all others are coupled by a coefficient $\kappa$ (Fig.~\ref{fig:genevol}b).
The first waveguide is also characterized by a propagation-constant detuning $\varepsilon$, which, from the quantum evolution point of view, corresponds to the energy detuning of the site; all other waveguides are identical.  As shown in Fig.~\ref{fig:genevol}c, such a system actually consists in a discrete state (the first site), coupled to a continuum band of width $4 \kappa$. \cite{nota1}

The dynamics of this quantum system is generated by the Hamiltonian:
\begin{equation}\label{Hamiltonian}
H=H_0+H_1+H_{\mathrm{int}},
\end{equation}
with
\begin{align}
H_0 & =  \varepsilon \ket{0}\bra{0} , \\
H_1 & = \kappa \sum_{n\geq 1} \left( \ket{n}\bra{n+1} + \ket{n+1}\bra{n} \right) \nonumber \\ & + q \sum_{n\geq 1} \left( \ket{n}\bra{n+2} + \ket{n+2}\bra{n} \right), \\
H_{\mathrm{int}} & =  \kappa_0 (\ket{0}\bra{1} + \ket{1}\bra{0}) + q_0 (\ket{0}\bra{2} + \ket{2}\bra{0}) .
\end{align}
Note that we have also included a next-nearest-neighbor hopping term, characterized by a coupling coefficient $q$ ($q_0$ for the first waveguide). This additional interaction is unavoidable in our experimental setting and its effects are typically small \cite{keil2015}, but can become quantitatively relevant in the high-depletion (long-time) regimes; we will assume for simplicity $q_0=q$ in the numerical simulations.

We consider the system initialized at $t=0$ in the first site of the array, $\ket{\psi_0}=\ket{0}$. The quantity typically chosen to investigate the temporal behaviour of the system is the survival probability, defined (with $\hbar=1$) as
\begin{equation}
p(t)=|\As(t)|^2, \qquad  \As(t) = \langle \psi_0 \ket{\psi(t)} = \bra{\psi_0} \ee^{-\ii t H} \ket{\psi_0}, 
\label{eq:evEquation}
\end{equation}
The initial state is \textit{unstable}, i.e. $p(t)\to 0$ as $t\to\infty$,  if $\lambda^2 < 1 - |\varepsilon|/2\kappa$ \cite{suppMat1}, being $\lambda = \kappa_0/\kappa$. The survival probability amplitude is the sum of two terms \cite{suppMat1}
\begin{equation}
\As(t) = \mathcal{Z} \ee^{-\ii E_P t} + \As_{\mathrm{cut}}(t) ,
\end{equation}
where $E_P$ is the pole of the propagator $G(z)=(z-H)^{-1}$ in the second Riemann sheet \cite{cohentannoudji}, whose imaginary component yields the decay rate. The exponential law is normally dominant at intermediate decay times. $\mathcal{Z}$ is called the \textit{wave-function renormalization}, and can be determined by extrapolating the exponential probability back to $t=0$:
\begin{equation}\label{eq:zeta}
Z=|\mathcal{Z}|^2 = 1 + \frac{\lambda^2}{1-\lambda^2} \frac{1- \frac{3}{4}\lambda^2 -\left(\frac{\varepsilon}{2\kappa}\right)^2}{1- \lambda^2 - \left(\frac{\varepsilon}{2\kappa}\right)^2} .
\end{equation}
It is possible to check that, whenever the imaginary part of $E_P$ is nonvanishing, the value of $Z$ provided by the above equation is strictly larger than one. The term $\As_{\mathrm{cut}}(t)$ accounts for all deviations from the exponential and dominates at short and long times. {In general, interference between the pole and cut terms also generates oscillations in the survival probability.}

The survival probability at short times can be extracted by a power-series expansion of the evolution operator $e^{-\ii t H}$, resulting in
\begin{equation}
p(t) = 1 - \left(\frac{t}{\tau_Z}\right)^2 + \Ord(t^4) .
\end{equation} 
The survival probability is thus quadratic at very short times, with curvature determined by the \textit{Zeno time} 
\begin{equation}\label{zt}
\tau_Z = \left( \bra{\psi_0} H^2 \ket{\psi_0} - \bra{\psi_0} H \ket{\psi_0}^2 \right)^{-1/2} = \frac{1}{\kappa_0} .
\end{equation}

At long times the contribution of $\As_{\mathrm{cut}}(t)$ becomes dominant over the exponential term, accounting for the power-law behaviour. In our case, the survival probability at long times reads
\begin{equation}\label{eq:powerlaw}
p(t)\simeq |a_{\mathrm{cut}}(t)|^2 = \left(\frac{C(t)}{t}\right)^3 (1+ \alpha(t) \cos (4 \kappa t + \varphi(t)) ),
\end{equation}
where $C(t)$, $\alpha(t)$ and $\varphi(t)$ become constant at sufficiently long times. {This result is to be expected from first principles \cite{Khalfin57,Khalfin58,Exner85}. An exponential behavior at all times would imply a Lorentzian energy density distribution, with support on all (positive and negative) energies, i.e.\ an unbounded Hamiltonian. Instead, if the spectrum is bounded from below, with a finite ground-state energy, the Paley-Wiener theorem states that the function $\left[\ln p(t)\right]/(1+t^2)$ is integrable, and thus $p(t)$ must be slower than exponential at long enough times.} In general, the power law appearing in \eqref{eq:powerlaw} is related to the structure of the coupling and the behavior of the density of states at energies close to the edge(s) of the continuum. {The oscillatory behavior, with angular frequency $4\kappa$, is due to the interference between the contributions from the two band edges.} Finally notice that the exact expression of the Zeno time \eqref{zt} is left unchanged by the introduction of next-to-nearest-neighbor couplings, as well as the form of the long-time survival probability \eqref{eq:powerlaw}, provided $q$ is real and $|q/\kappa|<1/4$.

In our experiments, waveguides were fabricated in fused silica substrate by femtosecond laser direct inscription. This technique exploits the nonlinear absorption of focused ultrashort laser pulses to induce permanent and localized refractive index modifications in the bulk of transparent dielectric materials. We used the second harmonic (520 nm wavelength) of an ytterbium femtosecond laser  (HighQ Spirit One), producing $\simeq$400 fs duration pulses at 20~kHz repetition rate. In our experiments, laser pulses with 350 nJ energy were focused, by means of a 0.45~NA microscope objective, 170~$\mu$m below the glass surface, and the substrate was translated with respect to the laser beam at constant speed comprised between 20 and 34~mm/s. Waveguides fabricated with these irradiation parameters yield single-mode behaviour for the 633~nm wavelength and propagation losses of 0.6 dB/cm.

To investigate the light propagation in waveguide arrays, we injected horizontally polarized light from a He:Ne laser source in the first waveguide (which corresponds to initializing the system in state $\ket{0}$) and imaged the structure from above, acquiring the scattered light at each point. The scattered signal is indeed locally proportional to the intensity of the propagating light.

As mentioned in the preceding analysis, we are mainly interested in retrieving the population of the first waveguide. This is not easy because such quantity spans a few orders of magnitude: at its entrance (i.e.\ at $t=0$), the first waveguide (namely state $\ket{0}$) is fully populated, but at later times, when the interesting power-law dynamics sets in, it may be heavily depleted. 

To perform the measurement, we developed a microscope assembly operating as a high-dynamic-range image scanner \cite{suppMat2}. The assembly is moved along the propagation coordinate by a computer-controlled motorized stage, with synchronized image acquisition by a CCD. To enhance the dynamic range of the measurement, pictures taken at different exposure times are combined and analyzed together.
The experimental survival probability $p(t)$ is retrieved as the ratio between the optical power scattered from the first waveguide and the global scattered power at each $t$. In this way, propagation losses of the waveguides, which are uniform in the array, are also normalized out and do not affect our results.

The coupling coefficients $\kappa$, $\kappa_0$ and $q$ depend on the relative distance between the waveguides, while the detuning $\varepsilon$ can be controlled by varying the writing speed \cite{crespiQuantumRabi}. These quantities have been calibrated in independent experiments where we fabricated several couples of identical parallel waveguides at different relative distances, and other couples of parallel waveguides that differ in writing speed. By observing the periodicity of the bouncing of coherent light  between the coupled waveguides \cite{yariv}, it is possible to measure the coupling coefficients and the propagation-constant detuning, thus retrieving their dependence on the inscription parameters.

We realized arrays characterized by different geometrical parameters, each containing 40 waveguides. To avoid boundary effects, we always chose coupling conditions in which light does not reach the last waveguide of the array within the propagation length ($<$~9~cm, size of our glass samples)\cite{suppInfFig3}. In such conditions the system dynamics is well explained by the semi-infinite model discussed above. Table~\ref{tab:arrays} displays the relevant physical parameters of three structures, which have been tailored in order to observe different dynamical regimes.

\begin{table*}[tb!]
\centering
\footnotesize
\begin{tabular}{c|c|c|c|c|c|c|c|c}
  & $d_0$ & $d$ & $v_0$ & $v$ & $\kappa_0$ & $\kappa$ & $\varepsilon$ & $q$\\
\hline \hline
\hspace{1em}A\hspace{1em} & 12.0~$\mu$m & 17.0~$\mu$m & 31.30~mm/s & 30.00~mm/s & 0.045~$\pm$0.001~mm$^{-1}$ & 0.119~$\pm$0.004~mm$^{-1}$ & -0.08~$\pm$~0.09~mm$^{-1}$ & 0.005~mm$^{-1}$\\
B & 15.5~$\mu$m & 15.0~$\mu$m & 30.00~mm/s & 31.30~mm/s & 0.118~$\pm$0.004~mm$^{-1}$ & 0.132~$\pm$0.004~mm$^{-1}$ & 0.10~$\pm$~0.09~mm$^{-1}$ & 0.01~mm$^{-1}$\\
C & 14.0~$\mu$m & 15.0~$\mu$m & 30.00~mm/s & 30.00~mm/s & 0.183~$\pm$0.006~mm$^{-1}$ & 0.158~$\pm$0.005~mm$^{-1}$ & 0.0~mm$^{-1}$ & 0.01~mm$^{-1}$\\
\end{tabular}
\caption{Relevant physical parameters of the three waveguide arrays (A, B, C). $d_0$ is the distance between the first and the second waveguide, $d$ the distance between all other neighbouring waveguides. $v_0$ is the writing speed of the first waveguide, $v$ the writing speed of all the other ones. $\kappa_0$, $\kappa$, $\varepsilon$ and $q$ have the same meaning as in Eq.~(\ref{eq:evEquation}); the reported values for each array are the nominal ones, estimated on the basis of the preliminary calibration experiments. Errors correspond to standard deviations and are due to tolerances in the waveguide inscription process; they are also estimated by means of the calibration experiments. Where not written explicitly, uncertainty is indicated by the number of significant digits used. }
\label{tab:arrays}
\end{table*}

\begin{figure}[t]
\centering
\includegraphics{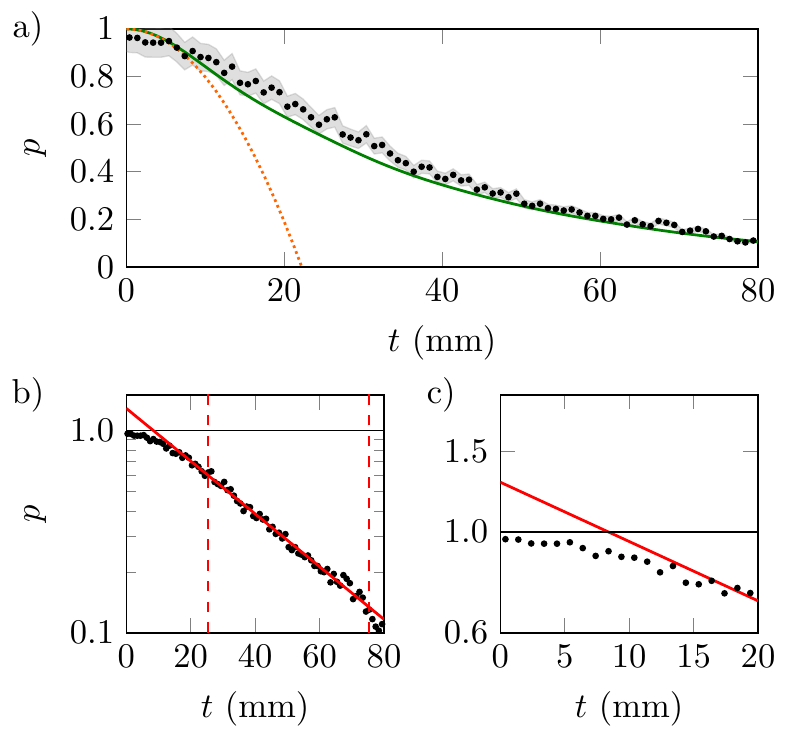}
\caption{Experimental survival probability in a weakly coupled system (array A in Table~\ref{tab:arrays}). Experimental points (black dots) are reported in linear (a) and semilogarithmic plots (b-c). Experimental errors are shown in the linear plot as a gray area around the points. The green continuous line in panel (a) is the theoretical prediction, obtained by solving Eq.~\eqref{eq:evEquation} with the nominal data ($\kappa_0$, $\kappa$, $\varepsilon$, $q$) of array A. The orange dotted line in panel (a) is the parabolic trend $1-(t/\tau_z)^2$ with $\tau_z=1/\kappa_0$. The red continuous line in panels (b) and (c) is a linear fit performed on the semilogarithmic plot, considering the data between the two dashed red lines in panel (b).}
\label{fig:case1}
\end{figure}

The transition between the initial Zeno region and the exponential decay is evident for systems in which state $\ket{0}$ is weakly coupled to the continuum. In Fig.~\ref{fig:case1}, we report the experimental decay of a system designed with $\lambda \simeq 0.37$, whose parameters are listed in line A of Table~\ref{tab:arrays}. Figure~\ref{fig:case1}a shows, in linear scale, the full evolution analyzed in the experiment. An initial quadratic region is manifest at early times. The subsequent exponential behavior is plainly revealed in Fig.~\ref{fig:case1}b, where the same data are plotted in a semi-logarithmic graph: a linear fit, corresponding to an exponential decay in linear scale, is also plotted. By zooming in the propagation region below 20~mm (Fig.~\ref{fig:case1}c) one can see that the intercept of this straight line, corresponding to the wavefunction renormalization parameter $Z$ in Eq.~\eqref{eq:zeta}, falls above 1. Although the fitted value $Z\simeq 1.23$ is slightly larger than expected, the experimental outcomes confirm the theoretical prediction $Z>1$ for the analyzed dynamics, independent of the specific values of the parameters.

\begin{figure}[t]
\centering
\includegraphics{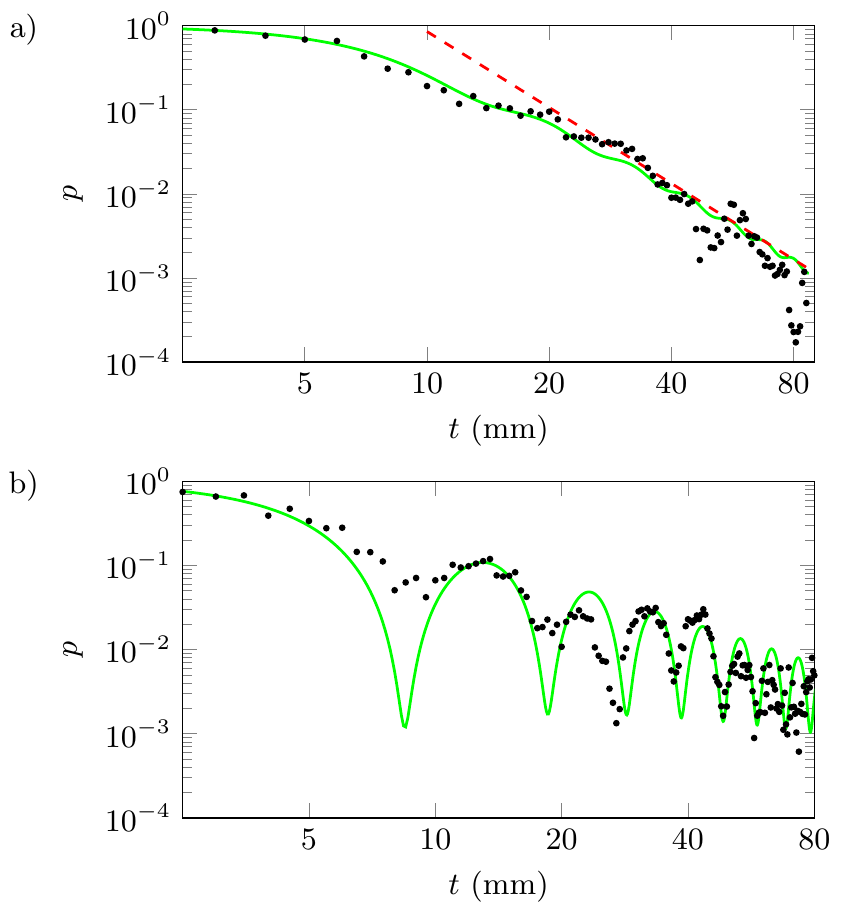}
\caption{Experimental survival probability in systems with stronger coupling, i.e. array B (panel (a)) and array C (panel (b)) in Table~\ref{tab:arrays}. Experimental points (black dots) are reported in logarithmic plots; experimental errors are comparable to the marker size. The (green) solid lines represent data fitting, obtained by solving Eq.~\eqref{eq:evEquation} using: (a) $\kappa_0$~=~0.119~mm$^{-1}$, $\kappa$~=~0.132~mm$^{-1}$, $\varepsilon$~=~0.12~mm$^{-1}$, $q$~=~0.01~mm$^{-1}$.  (b) $\kappa_0$~=~0.205~mm$^{-1}$, $\kappa$~=~0.160~mm$^{-1}$, $\varepsilon$~=~0~mm$^{-1}$, $q$~=~0.01~mm$^{-1}$; the fitting parameters are close to the nominal ones reported in Table I. The (red) dashed line in panel (a) represents the polynomial trend $p(t)\simeq (C_{\infty}/t)^3$.} 
\label{fig:case34}
\end{figure}

The large-time behavior predicted in Eq.~\eqref{eq:powerlaw}, consisting in a $t^{-3}$ power-law tail with superimposed oscillations, can be better appreciated in systems with stronger couplings. Figure~\ref{fig:case34} shows the measured decay for two systems, corresponding to cases B and C in Table~\ref{tab:arrays} and featuring $\lambda \simeq 0.89$ and  $\lambda\simeq 1.16$ respectively. In case C, the energy detuning of the system is zero (waveguides are written with the same propagation constant) and oscillations are more pronounced, while in case B, where the detuning is relevant ($\varepsilon\simeq \kappa$), oscillations are almost suppressed. 

In case B, the experimentally observed decay follows with good approximation the theoretical (solid green) line, which, at $t\simeq 40\,\mathrm{mm}$, relaxes towards the asymptotic power law $p(t)\simeq (C_{\infty}/t)^3$ (see Eq.\ \eqref{eq:powerlaw}), with $C_{\infty}=9.48\,\mathrm{mm}$. Thus, the dynamics of case B features a pure power-law behavior at times which are long, but still within reach of the experiment. It must be noted that the choice of the parameters has allowed the observation of the onset of such regime when the state was not heavily depleted yet ($p(t=40\,\mathrm{mm})\sim 10^{-2}$). Even for case C, theoretical simulations show that the power law takes place at sufficiently long times, but here such times are far beyond the experimental reach by an order of magnitude. Therefore, the behavior observed in Fig.~\ref{fig:case34}b, although sub-exponential, cannot be described by a single power law. The most interesting feature of the time evolution in case C is the presence of oscillations with period $\pi/(2\kappa)$ (see Eq.\ \eqref{eq:powerlaw}), which are due to the coupling with a bounded continuum, with bandwidth $4\kappa$. Such oscillations are entirely due to the cut contributions to the survival amplitude, and cannot be described by any Markovian approximation. 

We have reported the experimental observation of different decay regimes, typical of genuinely quantum dynamics, using photons propagating in waveguide arrays. The femtosecond laser writing technology allowed us to define with high control the relevant physical properties of the system. The imaging technique here developed enabled the measurement of the light distribution during propagation in the array with an unprecedented dynamic range.

Note that our characterization technique, which relies on the scattered light, differs from that used in other experiments in the literature, where fluorescence emission from the waveguides was exploited \cite{longhiRev,szameitNolte,crespiQuantumRabi}. In addition, here we adopted a multi-exposure acquisition technique to extend the available dynamic range beyond the 8-bit limit of the camera, allowing us to compare intensity levels which differ by a factor larger than $10^4$. In this way we have been able to characterize power-law decay tails, which are generally very elusive to experimental observation. We believe that these results open novel perspectives in the study of quantum decay dynamics, as well as in the investigation of the interaction between a system and its environment, including noise-enhanced transport phenomena or non-Markovian processes.

\begin{acknowledgments}
RO acknowledges financial support by the European Research Council (ERC) Advanced Grant CAPABLE (grant agreement no. 742745). PF and SP are partially supported by Istituto Nazionale di Fisica Nucleare (INFN) through the project ``QUANTUM". FVP is supported by INFN through the project ``PICS''. PF is partially supported by the Italian National Group of Mathematical Physics (GNFM-INdAM).

\end{acknowledgments}


\cleardoublepage
\onecolumngrid

\begin{center}
\large \bf - Supplemental Material - 
\end{center}
\appendix
\section{I. Theoretical model}

We will consider a physical system that can be efficiently simulated and controlled in an experimental setup, on the basis of the quantum-optical analogy. The system consists of a linear semi-infinite array of sites, corresponding to states $\ket{n}$, with $n\geq0$. The system is initialized at $t=0$ in the first site of the array, $\ket{\psi_0}=\ket{0}$, and its dynamics is generated by the Hamiltonian 
\begin{equation}
H=H_0+H_1+H_{\mathrm{int}},
\end{equation}
with
\begin{align}
H_0 & =  \varepsilon \ket{0}\bra{0} , \\
H_1 & = \kappa \sum_{n\geq 1} \left( \ket{n}\bra{n+1} + \ket{n+1}\bra{n} \right) \nonumber + q \sum_{n\geq 1} \left( \ket{n}\bra{n+2} + \ket{n+2}\bra{n} \right), \\
H_{\mathrm{int}} & =  \kappa_0 (\ket{0}\bra{1} + \ket{1}\bra{0}) + q_0 (\ket{0}\bra{2} + \ket{2}\bra{0}) .
\end{align}
The next-to-nearest-neighbor hopping has been included since it is unavoidable in the experimental setting, in which $q/\kappa\simeq 0.17$. However, we shall initally consider the case $q_0=q=0$ to highlight the relevant physics of the model and the transition between different regimes in the time evolution. 

If $q=0$, the Hamiltonian $H_1$ is exactly diagonalizable as
\begin{align}
H_1 & = 2\kappa \int_0^{\pi} \mathrm{d}k  \cos k \, \ket{\varphi(k)}\bra{\varphi(k)} , \\
\ket{\varphi(k)} & = \sqrt{\frac{2}{\pi}} \sum_{n\geq1} \sin(kn) \,\ket{n} .
\end{align}
In general, the survival amplitude can be determined by a Fourier-Laplace transform
\begin{equation}\label{temporal}
\As(t) = \frac{\ii}{2\pi} \int_{-\infty+\ii 0^+}^{+\infty+\ii 0^+} \mathrm{d} E \,\ee^{-\ii Et} G(E),
\end{equation}
where the propagator of the initial state in the energy domain reads
\begin{equation}
G(E)= \bra{\psi_0}\frac{1}{E-H}\ket{\psi_0} .
\end{equation}

\begin{figure}
\centering
\includegraphics[width=0.5\textwidth]{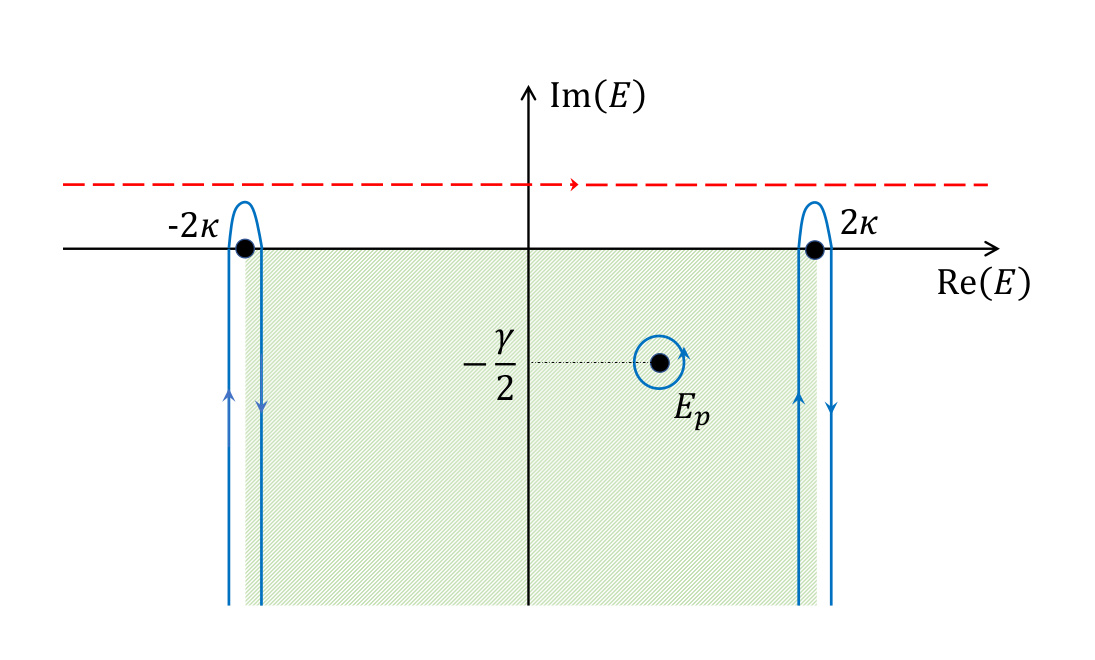}
\caption{Representation of the  integration paths used to obtain the survival amplitude from its Fourier-Laplace transform $G(E)$. The (red) dashed line above the real axis represents the  integration path in Eq.~\eqref{temporal}, that can be deformed into the (blue) solid path, composed of three separate curves. Part of the new path lies on the second Riemann sheet of the propagator (shaded area), where $G(E)$ has been analytically continued from above across the cut singularity in $[-2\kappa,2\kappa]$. The imaginary part of the pole $E_p$ determines the decay rate $\gamma$.}\label{fig:path}
\end{figure}

The propagator can be analitically determined and reads ($\lambda= \kappa_0/\kappa$)
\begin{equation}\label{propagator}
G(E) = \frac{1}{ E- \varepsilon - \frac{\lambda^2}{2} E- \frac{\lambda^2}{2} \sqrt{E+2\kappa}\sqrt{E-2\kappa}  } ,
\end{equation}
where the principal square root $\sqrt{z}=\sqrt{|z|}\ee^{\ii \mathrm{Arg}(z)/2}$ is assumed, with $\mathrm{Arg}(z)\in(-\pi,\pi]$. The propagator, which must  be analytic in the whole complex plane out of the real axis, is characterized for all values of the physical parameters by a cut singularity, due to a jump in the imaginary part of the product of square roots, on the segment $E\in(-2\kappa,2\kappa)$, corresponding to the continuum spectrum of $H_1$. A rigorous way to separate the exponential contribution to the survival amplitude from the deviations requires to perform an analytic continuation of the propagator below the cut, and to deform the integration path in \eqref{temporal} into the one in Fig.\ \ref{fig:path}, which partly lies on the second Riemann sheet, where the analytic continuation $G^{\mathrm{II}}(E)$ is characterized by a ``$+$'' sign in front of the square roots in \eqref{propagator}.

The expression of the only pole $E_p$ of the analytic continuation that lies in the lower half-plane, satisfying $(G^{\mathrm{II}}(E_p))^{-1}=0$ with $\mathrm{Im}(E_p)<0$, reads
\begin{equation}\label{Epole}
E_p = \frac{1}{1-\lambda^2} \left[ \left(1-\frac{\lambda^2}{2}\right) \varepsilon -\ii \lambda^2 \kappa \sqrt{1-\lambda^2-\left(\frac{\varepsilon}{2\kappa}\right)^2} \right]
\end{equation}
and lies between the vertical half-lines $(-2\kappa-\ii\infty,-2\kappa)$ and $(2\kappa-\ii\infty,2\kappa)$ for $\lambda^2/2<(2\kappa-|\varepsilon|)/(4\kappa-|\varepsilon|)$. In this case, the survival amplitude $\As$ 
can be written
\begin{equation}
\As(t) = \bra{\psi_0} \ee^{-\ii t H} \ket{\psi_0} = \mathcal{Z} \ee^{-\ii \Delta t - \frac{\gamma}{2} t} + \As_{\mathrm{cut}}(t) ,
\end{equation}
where $\Delta=\mathrm{Re}(E_p)$, $\gamma=-2\mathrm{\Im}(E_p)$, and $\mathcal{Z}$, called the \textit{wave function renormalization}, is the residue of $G^{\mathrm{II}}$ at $E_p$, from which one can obtain the extrapolated value of the exponential probability back at $t=0$
\begin{equation}
Z=|\mathcal{Z}|^2 = 1 + \frac{\lambda^2}{1-\lambda^2} \frac{1- \frac{3}{4}\lambda^2 -\left(\frac{\varepsilon}{2\kappa}\right)^2}{1- \lambda^2 - \left(\frac{\varepsilon}{2\kappa}\right)^2} ,
\end{equation}
which turns out to be strictly larger than unity and increasing with both $\lambda$ and $|\varepsilon|$ whenever $E_p$ has nonvanishing imaginary part. The decay rate $\gamma$ can be evaluated through the Fermi golden rule:
\begin{equation}
\gamma = 2\pi \int_0^{\pi} \mathrm{d}k\, |\bra{0}H_{\mathrm{int}}\ket{\varphi(k)}|^2 \delta (2\kappa\cos k-\varepsilon) = 2 \lambda^2 \kappa \sqrt{1-\left(\frac{\varepsilon}{2\kappa}\right)^2}.
\end{equation}
The cut contribution
\begin{equation}\label{Acut}
\As_{\mathrm{cut}}(t) = - \frac{\lambda^2}{2\pi} \sum_{\sigma=\pm} \ee^{-\ii \sigma\!\left(2\kappa t+\frac{\pi}{4}\right)} \int_0^{\infty} \mathrm{d}x\, \ee^{-xt} \sqrt{x} \, \mathcal{C}_{\sigma}(x) , 
\end{equation}
with
\begin{equation}
\mathcal{C}_{\sigma}(x) = \frac{\sqrt{4\kappa-\ii\sigma x}}{ \left[\left(1-\frac{\lambda^2}{2}\right)(2\sigma\kappa-\ii x) - \varepsilon\right]^2 +\frac{\lambda^4}{4} (4\ii\sigma\kappa +x)x} ,
\end{equation}
accounts for all deviations from the exponential behavior, in particular those at short times (quadratic Zeno region) and long times (power-law tail). The Zeno time can be calculated from the short-time expansion of the evolution, yielding
\begin{equation}
\tau_Z = \frac{1}{\kappa_0} = \frac{1}{\lambda \kappa} .
\end{equation}
The expression \eqref{Acut} is instead the most useful tool to determine the behavior at long times, in which the small values of the integration variable (namely, the energies close to the branching points $\pm 2\kappa$) are relevant. The dominant contribution at long times, obtained by approximating $\mathcal{C}_{\sigma}(x)$ with $\mathcal{C}_{\sigma}(0)$ under the integral, yields the power-law survival probability [compare with Eq.~(10) of the main text]
\begin{equation}\label{power}
p(t) \simeq \frac{\lambda^4}{16\pi t^3} \left[ \mathcal{C}_+^2(0) + \mathcal{C}_-^2(0) + 2 \mathcal{C}_+(0)\mathcal{C}_-(0) \sin(4\kappa t) \right] .
\end{equation} 
Oscillations in \eqref{power} are due to the existence of an upper bound to the cut in the propagator (i.e.\, to the support of the spectral density) of the initial state, while the $t^{-3}$ behavior is evidently related, for dimensional reasons, to the square-root behavior of the spectral density close to the branching points. 

We can now define the transition times between the three regimes in the evolution of our system. The transition time $\tau_0$ between the quadratic approximation and the exponential can be chosen at the intersection of the two curves,
$\exp(-\tau_0^2/\tau_Z^2) = Z \exp(-\gamma\tau_0)$.
However, since $Z$ is strictly larger than one in our case, it is possible that such intersection does not exist, as it occurs for very small $\lambda$. Thus, $\tau_0$ can be more conveniently defined as the time of closest approach between the two curves, i.e.\
\begin{equation}
\tau_0 =\frac{\gamma \tau_Z^2}{2},
\end{equation}
which, for small values of $\lambda$, reads
\begin{equation}
\tau_0 = \frac{1}{\kappa}\sqrt{1-\left(\frac{\varepsilon}{2\kappa}\right)^2}.
\end{equation}

As for the second transition time $\tau_{\infty}$ between the exponential and power-law regimes, it can be defined as the intersection time between the damped exponential and the non-oscillating part of the power law, namely
\begin{equation}\label{tinf}
Z \exp(-\gamma \tau_{\infty}) = \frac{C_{\infty}}{\tau_{\infty}^3} ,
\end{equation}
with
\begin{equation}
C_{\infty} = \frac{\lambda^4 \kappa}{2\pi} \left( \frac{1}{[\kappa(2-\lambda^2)+\varepsilon]^4} + \frac{1}{[\kappa(2-\lambda^2)-\varepsilon]^4} \right) .
\end{equation}
For small values of $\lambda$ it diverges as
\begin{equation}
\tau_{\infty}\sim \frac{1}{\gamma} \log\frac{Z}{\gamma^3C_\infty} = O\left(\frac{1}{\lambda^2} \log\frac{1}{\lambda}\right).
\end{equation}

The prediction of experimental results is improved by including the next-to-nearest-neighbor terms in the Hamiltonian, with 
\begin{equation}
Q=\frac{q}{\kappa}=\frac{q_0}{\kappa_0} \simeq 0.17.
\end{equation} 
Inclusion of such terms makes the theory more complicated, though it can still be tackled with analytical methods. One finds that, for $0<Q<1/4$, the qualitative features of time evolution do not change with respect to the case $Q=0$.


\section{II. Experimental details}

\begin{figure}[h]
\centering
\includegraphics[height=8cm]{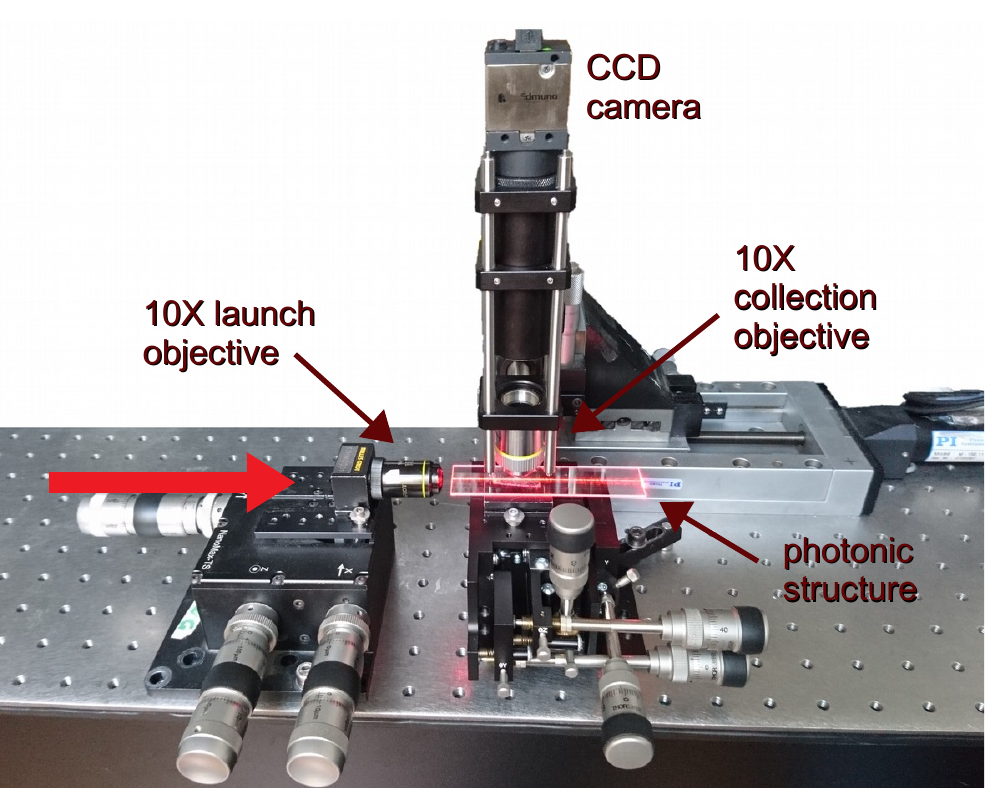}
\caption{\label{fig:experimental} Picture of the experimental apparatus employed to characterize the light distribution along propagation, in the fabricated waveguide arrays.}
\end{figure}
\begin{figure}[h]
\centering
\includegraphics[width=12cm]{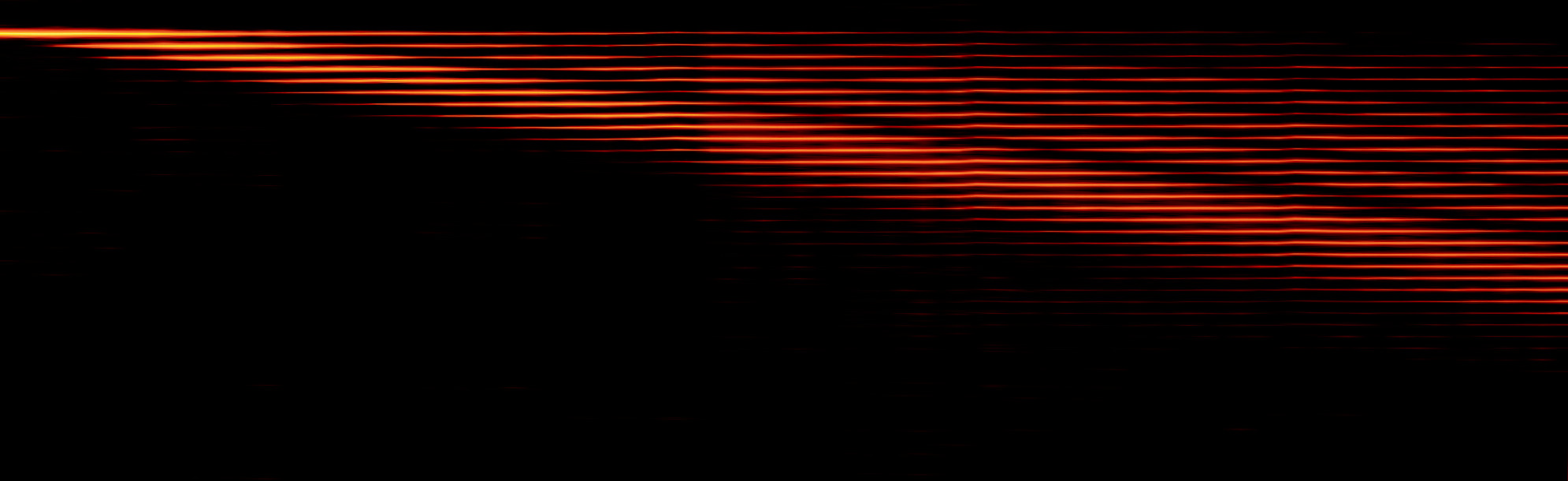}
\caption{\label{fig:array} Example of acquired intensity distribution from a waveguide array (array B in Table~1 of the Main Text). Laser light is coupled in the array from the left in the topmost waveguide. The original acquisition is monochrome and measures 89 (horizontal) $\times$ 1280 (vertical) pixels (corresponding to a physical region of about 88~mm $\times$ 640~$\mu$m sampled longitudinally with steps of 0.5~mm). Here the image is reshaped for better readability and is shown using a logarithmic false-color scale.
Note that the last illuminated waveguide is the 30$^\text{th}$ (counting from top) of the 40-waveguide
array. In this condition the semi-infinite model is valid.}
\end{figure}

\subsubsection{Details of the characterization apparatus}

Figure~\ref{fig:experimental} shows the experimental apparatus that we employed for the characterization of the waveguide arrays. The glass chip containing the photonic structures is fixed onto a 4-axis manual micrometric manipulator (Thorlabs MBT402D/M). A coherent-light beam from a He:Ne laser is focused on the entrance facet of the chip by a 10X objective. This laser is capable of emitting up to 15 mW power, but in our experiments it was attenuated to a few mW level. The chip is micrometrically aligned to the laser focus so that light is coupled only to the input of the first waveguide of the desired array. To facilitate this condition, only the first waveguide of each array reaches the input facet of the chip, while the other ones are fabricated starting about 10~mm inside. 

To acquire a picture from above of the light scattered from the waveguides, we employ a microscope-assembly composed of another 10X objective (NA~=~0.25) and an 8-bit CCD camera (Edmund Optics EO-1312M), mounted vertically on translation stages. The relative distance between the objective and the camera gives a magnification factor $\sim$10. Transversally, the position of the assembly is adjusted by manual micrometers; longitudinally, it is controlled by a motorized linear stage (P.I. M155-11) with precision better than 20~$\mu$m and travel range of 100~mm. The motion of the linear stage and the image acquisition from the CCD are synchronized via a MATLAB\textsuperscript{\textregistered} script, to operate similarly to an image scanner.

The assembly is translated with fixed steps along the light-propagation coordinate {and, at each step, several pictures are acquired, with different exposure times (ranging from about 1~ms to about 63~ms). 
Each picture is integrated in a window of width $w$ along the propagation coordinate to average the speckle noise, and give a reliable measurement of the intensity distribution along the transverse coordinate. Such integration window corresponded to 400~$\mu$m for the experiment in Figure~2 of the Main Text, and 200~$\mu$m in the other cases. In addition, the intensity values are normalized, for each picture, to the exposure time: this allows one to obtain measurements that are consistent and comparable one to the other.

First, we reconstruct an image of the whole array (Fig.~\ref{fig:array}) where the exposure times of the different sections are chosen in such a way that the highest overall signal is provided, but avoiding saturation in any point.  Then, we focus on the first waveguide and we assemble another image of the whole array, in which, on the contrary, we choose for each section the longest exposure time that avoids saturation only in the first waveguide.

The first image is then integrated transversally over all the waveguides, to obtain the overall power propagating in the array at each propagation coordinate $P_{\mathrm{tot}}(t)$. The second image is also integrated transversally, but including only the pixels of the first waveguide: this enables one to measure the power propagating in the first waveguide $P_1 (t)$. Finally, the experimental survival probability is retrieved as $p(t) = P_1(t)/P_{\mathrm{tot}}(t)$. It should be noted that both $P_1 (t)$ and $P_{\mathrm{tot}} (t)$ are affected by the propagation losses in the waveguide array. However, being this effect the same for the two quantities, it cancels out in the ratio that defines $p(t)$.}

Note that, combining the 2$^8$~=~256 available dynamic levels of the camera with exposures varying by a factor of about 2$^6$=64, we can compare intensity levels which differ by a factor of about $2^{14} > 10^4$, avoiding saturation in any point of the image. On the other hand, the intensity measurements on each pixel are still (properly rescaled) data acquisitions made with the 256-level camera. Therefore the available dynamic range is about 14 bits, while the precision in evaluating the intensity of each pixel is 8 bits.


\subsubsection{Estimate of uncertainties}

We describe in the following how we estimated the uncertainties for the physical quantities involved in the experimental measurements. 
\begin{itemize}
\item The uncertainty in the propagation coordinate $t$ derives from the finite (and not infinitesimal) width of the spatial integration window in the acquisition procedure. We estimate the uncertainty value as the standard deviation associated to a uniform probability distribution with the same width $w$ as the integration window. 
\item The uncertainty in the measured values of the survival probability is estimated by assuming that the main contribution to measurement noise is due to speckle and non-uniformity of the scattering centers. 
To estimate quantitatively the noise we start by considering the measured value of total optical power propagating in the array at each point $t$ (as inferred from the acquired images of the array). In the ideal case the total optical power should follow an exponential decay trend; we thus fit an exponential curve on its values as a function of $t$. The deviation of these experimental point from the fit is attributed to the speckle and non-uniformities of the scattering centers (we also note that such noise is proportional to the signal amplitude). We use the standard deviation of the residuals of the fit, normalized to the fitted point values, as an estimation of $\sigma_p/p$. We can thus associate to each point $p(t)$ an error bar with amplitude $\sigma(t) = p(t)\cdot \sigma_p/p$.
\item The uncertainty in the values of the coupling coefficients ($\kappa_0$, $\kappa$, $q$) and of the propagation constant detuning ($q$) is assumed to be entirely due to non-reproducibility of the coupling coefficients within the same fabrication session. This uncertainty was evaluated from the preliminary experiments, performed to calibrate their dependency on the inscription parameters.
\end{itemize}


\begin{thebibliography}{10}
\bibitem{nota1} This is true also in the case in which next-to-nearest-neighbor waveguides are coupled, with coupling constant $q$ (see Eq.~\eqref{Hamiltonian}), provided that $|q/\kappa|<1/4$.
\end{thebibliography}
\end{document}